\documentclass[Afour,sageh,times]{sagej}
\usepackage{moreverb,url}

\newcommand\BibTeX{{\rmfamily B\kern-.05em \textsc{i\kern-.025em b}\kern-.08em
T\kern-.1667em\lower.7ex\hbox{E}\kern-.125emX}}

\def\volumeyear{2015}
\usepackage{amsfonts}
\usepackage{epstopdf}
\usepackage{bm}
\usepackage{booktabs}

\def\b{{\bf b}}

\def\X{{\bf X}}

\def\X{{\bf X}}

\def\b{{\bf b}}

\def\sig0{\sigma_0}
\def\sig1{\sigma_1}

\def\bu{({\bf{b}},{\bf{u}})}
\def\bu0{({\bf{b}}_0,{\bf{u}}_0)}
\def\bu1{({\bf{b}}_1,{\bf{u}}_1)}

\def\but0{(\tilde{{\bf{b}}}_0,\tilde{{\bf{u}}}_0)}

\def\but{(\tilde{{\bf{b}}},\tilde{{\bf{u}}})}

\def\b{{\bf b}}

\begin{document}

\runninghead{Imani et al.}

\title{Sequential Experimental Design for Optimal Structural Intervention in Gene Regulatory Networks Based on the Mean Objective Cost of Uncertainty}

\author{Mahdi Imani\affilnum{1}, Roozbeh
Dehghannasiri\affilnum{2},  Ulisses M. Braga-Neto\affilnum{1,3} and Edward R. Dougherty\affilnum{1,3}}

\affiliation{\affilnum{1}Department of Electrical and Computer Engineering, Texas A\&M University, College Station, TX, 77843, USA\\
\affilnum{2}School of Medicine, Stanford University, Stanford, CA 94305, USA \\
\affilnum{3}Center for bioinformatics and Genomics Systems Engineering, TEES, College Station, TX 77845, USA.
}

\corrauth{Mahdi Imani}

\email{m.imani88@tamu.edu}

\begin{abstract}
Scientists are attempting to use models of ever increasing complexity, especially in medicine,
where gene-based diseases such as cancer require better modeling of cell regulation. Complex
models suffer from uncertainty and experiments are needed to reduce this uncertainty. Because
experiments can be costly and time-consuming it is desirable to determine experiments providing
the most useful information. If a sequence of experiments is to be performed, experimental
design is needed to determine the order. A classical approach is to maximally reduce the overall
uncertainty in the model, meaning maximal entropy reduction. A recently proposed method takes
into account both model uncertainty and the translational objective, for instance, optimal
structural intervention in gene regulatory networks, where the aim is to alter the regulatory logic
to maximally reduce the long-run likelihood of being in a cancerous state. The mean objective
cost of uncertainty (MOCU) quantifies uncertainty based on the degree to which model
uncertainty affects the objective. Experimental design involves choosing the experiment that
yields the greatest reduction in MOCU. This paper introduces finite-horizon dynamic
programming for MOCU-based sequential experimental design and compares it to the greedy
approach, which selects one experiment at a time without consideration of the full horizon of
experiments. A salient aspect of the paper is that it demonstrates the advantage of MOCU-based
design over the widely used entropy-based design for both greedy and dynamic-programming
strategies and investigates the effect of model conditions on the comparative performances.
\end{abstract}


\maketitle

\section{Introduction}

A basic problem in genomic signal processing is to derive intervention
strategies for gene regulatory networks (GRNs) to avoid undesirable states,
in particular, cancerous phenotypes. The problem goes back to the early days
of genomics when two paradigms were introduced to force dynamical gene
regulatory networks away from carcinogenic states, \textit{dynamical
Intervention}~(\citealp{shmulevich2002probabilistic,imani2018b,imani2018c,imani2018d}) and \textit{%
structural Intervention}~(\citealp{shmulevich2002control}). Substantial work
has been done since then [see \citealp{dougherty2010stationary} for reviews]. The goal of dynamical
intervention is to find an optimal finite or infinite-horizon control
strategy for altering the regulatory output of one or more genes at each
time point. The goal in structural intervention, which is the focus of the
current paper, is to find a one-time change in regulatory function for
beneficially changing the steady-state. The goal of dynamical intervention is
to find a proper finite or infinite-horizon control strategy for altering
the regulatory output of one or more genes at each time point. The goal of
structural intervention, which is the focus of the current paper, is to find
a one-time change in regulatory function for beneficially changing the
steady-state distribution of a GRN. A solution for optimal structure
intervention via representation of logical alterations of the regulatory
functions is discussed in~(\citealp{xiao2007impact}) in the context of
probabilistic Boolean networks (PBNs)~(\citealp{shmulevich2002boolean}). A
solution that applies Markov-chain perturbation theory to Markovian
regulatory networks to find a structural intervention that optimally reduces
the steady-state mass of undesirable states is presented in~(%
\citealp{qian2008effect}).

The basic theory of structural intervention provides optimal intervention
under the assumption that the regulatory model is known; however, in
practice, this is generally not the case. For instance, in a Boolean
network, or more generally a PBN, it is commonly the case that certain
regulatory relations are unknown, or at least not known with certainty. In
this case, one needs to reformulate the optimization problem to take into
account the uncertainty. While we are focusing here on gene regulatory
networks, this is a problem recognized as far back as the 1960s in control
theory (\citealp{bellman1959dynamic,Silver,Gozzolino}). More recently, it has been treated in signal processing, first
in a minimax framework (\citealp{kuznetsov1976stable,Kassam,poor1980robust}) and then in a Bayesian framework (\citealp{grigoryan2001bayesian,dalton2014}), and it
has also studied in pattern recognition (\citealp{dalton2013a,dalton2013b}). In the case of gene
regulatory networks, the problem has been addressed in (\citealp{yoon2013quantifying}) by utilizing
the mean objective cost of uncertainty (MOCU), which quantifies the
uncertainty based on its effect on the objective, in this instance, the
degree to which the uncertainty reduces the phenotypical effect of the
intervention.

In all cases, one would like to reduce the uncertainty in the model system
to achieve better optimal performance. Owing to cost and time, it is prudent
to prioritize potential experiments based on the information they provide
and then conduct the most informative. This process is called \emph{%
experimental design}. Various methods employ entropy~(%
\citealp{lindley1956measure,raiffa1961applied,huan2016sequential} 
), the mean objective cost of uncertainty (MOCU)~(%
\citealp{dehghannasiri2015optimal,dehghannasiri2015efficient,mohsenizadeh2016optimal}%
), and the knowledge gradient (KG)~(\citealp{frazier2008knowledge}). This
paper provides a comparison of entropy-based and MOCU-based methods.

Since uncertainty can be quantified via entropy, a historical approach to
experimental design has been to choose an experiment that maximally reduces
entropy~(\citealp{lindley1956measure,raiffa1961applied}). Assuming that the
true model lies in an uncertainty class of models governed by a probability
distribution, from a class of potential experiments the aim is to choose the
one that provides model information resulting in the greatest reduction of
entropy relative to the distribution. In the case of a GRN, uncertainty
might relate to lack of knowledge concerning regulatory relations and the
uncertainty class would consist of a collection of GRNs with differing
regulatory relations among some of the genes. Potential experiments would
characterize unknown regulatory relations, thereby reducing the uncertainty
class and lowering entropy.

Entropy does not take into account the objective for building a model. An
experiment might reduce entropy but have little or no effect on knowledge
necessary for accomplishing the desired objective. In the case of GRNs,
structural intervention involves altering gene regulation so as to reduce
the steady-state probability of undesirable states, such as
cell-proliferative (carcinogenic) states. Given a model, one finds an
optimal structural intervention~(\citealp{qian2008effect}). When there is
model uncertainty, we desire to reduce uncertainty relevant to determining
an optimal structural intervention. The mean objective cost of uncertainty~(%
\citealp{yoon2013quantifying}), which provides a quantification of
uncertainty based on the degree to which model uncertainty affects the
translational objective, is used for experimental design: choose the
experiment that yields the greatest reduction in MOCU~(%
\citealp{dehghannasiri2015optimal}).

Typically, one might perform a sequence of experiments to progressively
reduce the uncertainty. What should be the order of the experiments to get
the best reduction in uncertainty? Using MOCU, this problem has been
addressed in the context of gene regulatory networks in a greedy sequential
fashion (\citealp{yoon2013quantifying}): at each step the optimal experiment is chosen from among
the ones not yet performed. In the present paper, we demonstrate the
strength of MOCU-based experimental design by considering optimal sequential
experimental design in the context of structural intervention in Boolean
networks with perturbation (BNp). In this framework, there are $M$ unknown
regulatory relations and the aim is to sequentially choose optimal
experiments to reduce uncertainty. We compare design via MOCU and entropy
for both greedy and dynamic-programming-based sequential design, which has
not been previously used with MOCU.\vspace{-0ex}

\section{Gene Regulatory Networks and Interventions}

\subsection{Boolean Networks: A Brief Overview}

Several Boolean network models have been developed in recent years for
studying the dynamics of GRNs~(e.g. %
\citealp{kauffman1993origins,shmulevich2002probabilistic,imani2017,imani2018a}), for instance,
the cell cycle in the Drosophila fruit fly (\citealp{albert2003topology}),
in the Saccharomyces cerevisiae yeast (\citealp{kauffman2003random}), and
the mammalian cell cycle (\citealp{faure2006dynamical}).

A binary Boolean network (BN) on $n$ genes is represented by a set of
gene-expression, $\{X_{1},X_{2},...,X_{n}\},$ and a set of Boolean
functions, $\{f_{1},...,f_{n}\}$, that gives functional relationships
between the genes over time. The state of each gene is represented by 0
(OFF) or 1 (ON), where $X_{i}=1$ and $X_{i}=0$ correspond to the activation
and inactivation of gene $i$, respectively. The states of genes at time step 
$k$ is denoted by a vector $\X(k)=(X_{1}(k),...,X_{n}(k))$. The value of the $%
i$th gene at time step $k+1$ is affected by the value of the $k_{i}$
predictor genes at time step $k$ via $%
X_{i}(k+1)=f_{i}(X_{i_{1}}(k),X_{i_{2}}(k),...,X_{i_{k_{i}}}(k))$, for $%
i=1,...,n$. In a Boolean Network with perturbation (BNp), the state value of
each gene at each time point is assumed to be flipped with a small
probability $p$. This produces a dynamical model $\X(k+1)=\mathbf{F}%
(\X(k))\oplus {\mathbf{\eta}}(k)$, where $\mathbf{F}=(f_{1},...,f_{n})$, $%
\oplus $ is component-wise modulo 2 addition, and ${\mathbf{\eta}}%
(k)=(\eta_{1}(k),\eta_{2}(k),... ,\eta_{n}(k))$ with $\eta_{i}(k)\sim \text{Bernoulli}(p)$%
, for $i=1,...,n$. Letting $\{\mathbf{x}^{1},...,\mathbf{x}^{2^{n}}\}$ be
the set of all possible Boolean states, the transition probability matrix
(TPM) $\mathbf{P}$ of the Markov chain defined by the state model is given
by 
\begin{eqnarray}
{p}_{ij} &=&P\left( \X(k+1)=\mathbf{x}^{i}\mid \X(k)=\mathbf{x}^{j}\right) 
\nonumber \\
&=&p^{||\mathbf{x}^{i}\,\oplus \,\mathbf{F}(\mathbf{x}^{j})||_{1}}(1-p)^{n-||%
\mathbf{x}^{i}\,\oplus \,\mathbf{F}(\mathbf{x}^{j})||_{1}}\,,
\end{eqnarray}%
%
%
%
%
%
%
%
%
%
%
%
%
for $i,j=1,...,2^{n}$, where ${p}_{ij}$ refers to the element in the $i$th
row and $j$th column of the TPM matrix $\mathbf{P}$, and $||.||_{1}$ is the $%
L_{1}$ norm. For a non-zero perturbation process ($p>0$), the corresponding
Markov chain of a BNp possesses a steady-state distribution (SSD) $\pi $
describing the long-run behavior of system. The SSD can be computed based on
the TPM of the Markov chain as $\pi ^{T}=\pi ^{T}\mathbf{P}$, where $\mathbf{%
v}^{T}$ denotes the transpose of $\mathbf{v}$ and the $i$th element denotes
the steady-state probability of being at state $\mathbf{x}^{i}$.

\subsection{Structural Intervention of GRNs}

We now briefly review the solution that applies Markov-chain perturbation
theory to the TPM to find a structural intervention that optimally reduces
the steady-state mass of undesirable states~(\citealp{qian2008effect}).
Given a known BNp, under the rank-1 function perturbation the TPM $\mathbf{P}
$ will be altered to $\tilde{\mathbf{P}}=\mathbf{P}+\mathbf{a}\mathbf{b}^{T}$
(\citealp{qian2008effect}), where $\mathbf{a}$ and $\mathbf{b}$ are
arbitrary vectors and $\mathbf{b}^{T}\mathbf{e}=0$ ($\mathbf{e}$ is the all
unity column vector). A special case of a rank-1 perturbation, called a
single-gene perturbation process, is considered in this paper. According to
this process, the output state of a single input state changes and the
output states of other states stay unchanged. Let $\Psi $ be a class of
potential interventions to the network. Let $\tilde{\mathbf{F}}=(\tilde{f}%
_{1},...,\tilde{f}_{n})$ be the Boolean function after intervention. A
single-gene perturbation for the input state $j$ changes the output value of
the Boolean function: $s=\tilde{\mathbf{F}}(\mathbf{x}^{j})\neq \mathbf{F}(%
\mathbf{x}^{j})=r$, and $\tilde{\mathbf{F}}(\mathbf{x}^{i})=\mathbf{F}(%
\mathbf{x}^{i})$, for $i=1,...,2^{n}$ and $i\neq j$. We shall refer to this
as a $(j,s)$ intervention. The TPM after perturbation, $\tilde{\mathbf{P}%
\text{,}}$ is the same as $\mathbf{P}$, except for $\tilde{p}%
_{jr}=p_{jr}-(1-p)^{n}$ and $\tilde{p}_{js}=p_{js}+(1-p)^{n}$. The
steady-state distribution of the system after perturbation can be computed
as~(\citealp{qian2008effect}) 
\begin{equation}
\tilde{\pi}_{i}(j,s)=\pi _{i}+\frac{(1-p)^{n}\,\pi _{j}\,(z_{si}-z_{ri})}{%
1-(1-p)^{n}\,\,(z_{sj}-z_{rj})}\,,
\end{equation}%
where $\tilde{\pi}_{i}(j,s)$ is the steady state probability of the $i$th
state of the perturbed system following a $(j,s)$ intervention, $\mathbf{Z}=[%
\mathbf{I}-\mathbf{P}+\mathbf{e}\pi ^{T}]^{-1}$ is the fundamental matrix of
a BNp, $\mathbf{I}$ being the $n\times n$ identity matrix, and $%
z_{si},z_{ri},z_{sj},z_{rj}$ are elements of $\mathbf{Z}$.

If $U$ is the set of undesirable Boolean states, then $\tilde{\pi}%
_{U}(j,s)=\sum_{i\in U}\tilde{\pi}_{i}(j,s)$ is the steady-state probability
mass of undesirable states after applying a $(j,s)$ intervention. The
optimal single-gene perturbation structural intervention $(j^{\ast },s^{\ast
})$ minimizes $\tilde{\pi}_{U}(j,s)$: 
\begin{equation}
(j^{\ast },s^{\ast })=\operatornamewithlimits{argmin}_{_{j,s\in
\{1,2,...,2^{n}\}}}\tilde{\pi}_{U}(j,s)\,.  \label{OSI}
\end{equation}%
\vspace{0ex}%

\section{Experimental Design}

The complex regulatory machinery of the cell and the lack of sufficient data
for accurate inference create significant uncertainty in GRN models.
Consider a GRN possessing $M$ uncertain parameters $\theta ^{1},\theta
^{2},...,\theta ^{M}$. In our application, $\theta ^{i}$ corresponds to a
regulatory relation of an uncertain type that can take on two different
values: \textquotedblleft $\mathcal{A}$" for activating regulation and
\textquotedblleft $\mathcal{S}$" for suppressive regulation. These unknown
parameters result in $2^{M}$ different Boolean network models for the system
that differ in one or more of these uncertain regulations. Let $\Theta =\{%
\mathbf{\theta }_{1},...,\mathbf{\theta }_{2^{M}}\}$ be the uncertainty
class of these network models, where $\mathbf{\theta }_{j}\in \{\mathcal{A},%
\mathcal{S}\}^{M}$, for $j=1,...,2^{M}$. The prior distribution over Boolean
network models can be encoded into a single column vector 
\[
p(0)\,=\,\left[ P(\mathbf{\theta }^{\ast }=\mathbf{\theta }_{1}),...,P(%
\mathbf{\theta }^{\ast }=\mathbf{\theta }_{2^{M}})\right] ^{T}\,, 
\]%
%
%
%
%
%
%
%
%
%
%
%
where $\mathbf{\theta }^{\ast }$ is a vector containing the true values of
the parameters.

For a given initial distribution $p(0)$ and $i=1,...,M$, the prior
probability that the $i$th regulation is activating is 
\[
P(\theta ^{i}=\mathcal{A})\,=\,E_{p(0)}\left[1_{\mathbf{\theta }(i)=\mathcal{%
A}}\right]\,=\,\sum_{j=1}^{2^{M}}p_{j}(0)\,1_{\mathbf{\theta }_{j}(i)=%
\mathcal{A}}\,, 
\]%
where $1_{\mathbf{\theta }_{j}(i)=\mathcal{A}}=1$ if $\mathbf{\theta }{%
_{j}(i)=\mathcal{A}}$ and $0$ otherwise. The \textit{initial belief state}
is $b(0)=\left[ P(\theta ^{1}=\mathcal{A}),...,P(\theta ^{M}=\mathcal{A})%
\right] ^{T}$.

As in~\citealp{dehghannasiri2015optimal}, we assume that there exist $M$
experiments $T_{1},....,T_{M}$, where $T_{i}$ determines the regulation $%
\theta ^{i}$. More general experimental formulations are possible; for
instance, there is a probability that $T_{i}$ can return the wrong value (%
\citealp{mohsenizadeh2016optimal}).

Letting $\b(k)$ be the belief state before conducting the $k$th experiment.
Given that experiment $T_{i}$ at time step $k$ is performed, if the outcome
of the experiment shows that $\theta ^{i}=\mathcal{A}$, then the $i$th
element of the belief vector at time step $k+1$ will get the value $1$ and
the other elements will get their previous values: $\b_{i}(k+1)=1$, $%
\b_{l}(k+1)=\b_{l}(k)$ for $l=1,...,M,\,l\neq i$. On the other hand, if $%
\theta ^{i}=\mathcal{S}$, then the $i$th element of the belief vector will
be $0$ and the rest will be unaltered from time $k$ to $k+1$. 

Thus, each of $M$ elements of the belief vector can take three possible
values during the experimental design process and the belief vector is of
the form $\mathbb{B}=[\b^{1},...,\b^{3^{M}}]$, where $\b(k)\in {\mathbb{B}}$.
We view transition in the belief space as a Markov decision process (MDP)
with $3^{M}$ states. The \textit{controlled transition matrix} in the belief
space under experiment $T_{i}$ is a matrix of size $3^{M}\times 3^{M}$. The
element associated with the probability of transition from state $\b\in {%
\mathbb{B}}$ to state $\b^{\prime }\in {\mathbb{B}}$ under experiment $T_{i}$
can be written as 
\begin{eqnarray}
\mathbf{Tr}_{\b\b^{\prime }}({T_{i}}) &=&P\left( \b(k+1)=\b^{\prime }\mid
\b(k)=\b,T_{i}\right)  \nonumber  \label{eq-Tbb} \\
&=&\,\begin{cases} \b_i&\text{If }\b'_i=1\text{ and }\b_l=\b'_l \text{ for }
l\neq i,\\ 1-\b_i&\text{If }\b'_i=0\text{ and }\b_l=\b'_l \text{ for } l\neq
i,\\ 0 &\text{o.w. } \end{cases}
\end{eqnarray}

\subsection{Greedy MOCU}

Optimal experimental design using the \textit{mean objective cost of
uncertainty} (MOCU), first proposed in~\citealp{dehghannasiri2015optimal},
is briefly described in this section. Let $\xi _{\mathbf{\theta }}(\psi )$
be the cost of applying the intervention $\psi \in \Psi $ to the network $%
\mathbf{\theta }\in \Theta $. Using (\ref{OSI}), for any $\mathbf{\theta }%
\in \Theta $, the optimal single-gene perturbation structural intervention
for a BNp defined by a given uncertainty vector $\mathbf{\theta }$ is $\psi
_{\mathbf{\theta }}=(j_{\mathbf{\theta }}^{\ast },s_{\mathbf{\theta }}^{\ast
})$, where %
$\xi _{\mathbf{\theta }}(\psi )\geq \xi _{\mathbf{\theta }}(\psi _{\mathbf{%
\theta }})$ for any $\psi \in \Psi $.

The MOCU relative to an uncertainty class represented by the belief vector $%
\b $ and a class $\Psi $ of interventions is defined by 
\begin{eqnarray}
M_{\Psi }(\Theta\mid\b)\, &=&\,E_{\Theta\mid\b}\left[ \xi _{\mathbf{\theta }}(\psi _{\mathrm{IBR}%
}^{\theta\mid\b})-\xi _{\mathbf{\theta }}(\psi _{\mathbf{\theta }})\right] \,  \nonumber
\label{eq:MOCU} \\[-1ex]
&=&\,\sum_{j=1}^{2^{M}}p_{j}^{\b}\left[ \xi _{\mathbf{\theta }_{j}}(\psi _{%
\mathrm{IBR}}^{{\theta\mid\b}})-\xi _{\mathbf{\theta }_{j}}(\psi _{\mathbf{\theta }_{j}})%
\right] \,,
\end{eqnarray}%
where $\psi _{\mathrm{IBR}}^{\theta\mid\b}$ is an \textit{intrinsically Bayesian robust}
(IBR) intervention, 
\begin{equation}
\psi _{\mathrm{IBR}}^{\theta\mid\b}\,=\,\operatornamewithlimits{argmin}_{\psi \in \Psi
}E_{\b}[\xi _{\mathbf{\theta }}(\psi )]=\,\operatornamewithlimits{argmin}%
_{\psi \in \Psi }\sum_{j=1}^{2^{M}}p_{j}^{\b}\,\xi _{\mathbf{\theta }%
_{j}}(\psi ),\,
\end{equation}%
%
%
%
%
%
%
%
%
%
and $p^{\b}$ is the vector of posterior probabilities of network models for a
belief vector $\b$, which can be computed based on the independency of the regulations as 
\begin{equation}  \label{eq-pb}
p_{j}^{\b}\,=\,\prod_{i=1}^{M}\left[ \b_{i}1_{\mathbf{\theta }%
_{j}(i)=\mathcal{A}}+(1-\b_{i})1_{\mathbf{\theta }_{j}(i)=\mathcal{S}}\right] \,,
\end{equation}
for $j=1,...,2^{M}$. The IBR intervention $\psi _{\mathrm{IBR}}^{\theta\mid\b}$ depends
on the belief state $\b$, whereas the optimal intervention $\psi _{\mathbf{%
\theta }}$ is designed for a specific network model $\mathbf{\theta }\in
\Theta $. 
MOCU is the expected cost increase that results from applying a robust
intervention over all networks in $\mathbf{\theta }$ instead of the optimal
intervention for the unknown true network

The goal of sequential greedy MOCU-based experimental design (%
\citealp{dehghannasiri2015optimal}), referred to herein as Greedy-MOCU, is
to select an experiment at each time point that results in the maximal
reduction in MOCU in the next time step. If $\b$ is the current belief state,
then the Greedy-MOCU decision is given by 
\begin{eqnarray}
i^{\ast } &=&\operatornamewithlimits{argmin}_{i\in \{1,...,M\}}E_{\b^{\prime
}\mid \b,T_{i}}\left[ M_{\Psi }(\Theta\mid\b^{\prime })-M_{\Psi }(\Theta\mid\b)\right]  \nonumber \\
&=&\operatornamewithlimits{argmin}_{i\in \{1,...,M\}}\sum_{\b^{\prime }\in 
\mathbb{B}}\mathbf{Tr}_{\b\b^{\prime }}(T_{i})M_{\Psi }(\Theta\mid\b^{\prime })\,,
\end{eqnarray}%
where the second equality follows by expressing the expectation $%
E_{\b^{\prime }\mid \b,T_{i}}$ in terms of $\mathbf{Tr}_{\b\b^{\prime }}(T_{i})$%
, for $\b^{\prime }\in \mathbb{B}$, and then dropping the terms unrelated to
minimization. After making the decision and observing outcomes, one needs to
update the belief state and repeat another experimental design process if
necessary.

\subsection{Dynamic Programming MOCU}

Greedy-MOCU utilizes the expected value of MOCU in the next time step for
decision making at the current time. If the number of experiments, $N$, is
known a priori, then all future experiments (the remaining horizon) can be
taken into account during the decision making process. In this section, we
introduce optimal finite-horizon experimental design based on dynamic
programming (DP) (\citealp{bertsekas1995dynamic}), which we call DP-MOCU.

Let $\mu _{k}(\b)$ be a policy at time step $k$ that maps a belief vector $%
\b\in \mathbb{B}$ into an experiment in $\{T_{1},...,T_{M}\}$. We define a
bounded \emph{immediate cost function} at time step $k$ corresponding to
transition from the belief vector $\b(k)=\b$ into the belief vector $\b({k+1}%
)=\b^{\prime }$ under policy $\mu _{k}$ as 
\[
g_{k}\left( \b,\b^{\prime },\mu _{k}(\b)\right) \,=\,M_{\Psi }(\Theta\mid\b^{\prime
})\,-\,M_{\Psi }(\Theta\mid\b)\,, 
\]%
for $k=0,...,N-1$, where $g_{k}(\b,\b^{\prime },\mu _{k}(\b))\leq 0$. The \emph{%
terminal cost function} is defined as $g_{N}(\b)\,=\,M_{\Psi }(\Theta\mid\b)\,$, for any 
$\b\in \mathbb{B}$.

Letting $\Pi $ be the space of all possible policies, by using the
definitions of the immediate and terminal cost functions, an optimal policy, 
$\mu _{0:N-1}^{\mathrm{MOCU}}$, is given by solving the minimization problem 
\begin{eqnarray}  \label{eq:cost2}
&&\!\!\!\!\!\!\,\operatornamewithlimits{argmin}_{\mu _{0:N-1}\in \Pi }\!E\bigg[
\sum_{k=0}^{N-1}g_{k}\left( \b(k),\b(k+1),\mu _{k}(\b(k))\right) \nonumber\\
&&\qquad\quad\qquad\qquad\qquad\qquad\qquad+g_{N}(\b(N))%
\bigg],
\end{eqnarray}%
where the expectation is taken over stochasticities in belief transition.

Dynamic programming provides a solution for the minimization in (\ref%
{eq:cost2}). The method starts by setting the terminal cost function as $%
J_{N}^{\mathrm{MOCU}}(\b)\,=\,g_{N}(\b)\,,$for $\b\in \mathbb{B}$. Then, in a
recursively backward fashion, the optimal cost function can be computed as: 
\begin{eqnarray}
J_{k}^{\mathrm{MOCU}}(\b)&&=\min_{i\in \{1,...,M\}}\,E_{\b^{\prime }\mid
\b,T_{i}}\left[ g_{k}\left( \b,\b^{\prime },T_{i}\right)\right.\nonumber  \\
&&\qquad\quad\qquad\quad\qquad\qquad\left.+J_{k+1}^{\mathrm{MOCU}%
}(\b^{\prime })\right]  \nonumber \\
&&=\min_{i\in \{1,...,M\}}\!\!\left[ \sum_{\b^{\prime }\in \mathbb{B}}%
\mathbf{Tr}_{\b\b^{\prime }}({T_{i}})\left( g_{k}\left( \b,\b^{\prime
},T_{i}\right) \right.\right.\nonumber \\
&&\qquad\quad\quad\quad\qquad\qquad\left.\left.+J_{k+1}^{\mathrm{MOCU}}(\b^{\prime })\right) \right] ,
\end{eqnarray}%
with an optimal policy, $\mu _{k}^{\mathrm{MOCU}}(\b)$, given by 
\begin{equation}
\!\operatornamewithlimits{argmin}_{i\in \{1,...,M\}}\!\left[ \sum_{\b^{\prime
}\in \mathbb{B}}\mathbf{Tr}_{\b\b^{\prime }}({T_{i}})\left( g_{k}\left(
\b,\b^{\prime },T_{i}\right) +J_{k+1}^{\mathrm{MOCU}}(\b^{\prime })\right) %
\right] \!,
\end{equation}%
for $\b\in \mathbb{B}$ and $k=N-1,...,0$, where $\mathbf{Tr}({T_{i}})$ is
defined in (\ref{eq-Tbb}). Unlike Greedy-MOCU, the DP-MOCU policy decides
which uncertain regulation should be determined at each step in order to
maximally reduce the uncertainty relative to the objective after conducting
all $N$ experiments.

\subsection{Greedy Entropy}

The idea of entropy-based experimental design (%
\citealp{lindley1956measure,raiffa1961applied}) is to reduce the amount of
the entropy, which quantifies the uncertainty of the system. While
MOCU-based techniques take action to reduce the uncertainty with respect to
an objective, entropy-based techniques do not take into account the
objective during decision making. Performance comparisons are made in
Section~\ref{sec:NE}.

The entropy for belief vector $\b$ is $H(\b)\,=\,-\sum_{j=1}^{2^{M}}p_{j}^{\b}%
\,\log _{2}p_{j}^{\b}$, where $p^{\b}$ is the posterior probability of network
models under the belief state $\b$ defined in (\ref{eq-pb}). The maximum
value of the entropy is $M$, which corresponds to a uniform prior over the
network models, and the minimum value is $0,$ which corresponds to certainty.

The Greedy-Entropy approach sequentially chooses an experiment to minimize
the expected entropy at the next time step: 
\begin{eqnarray}
i^{\ast }\, &=&\,\operatornamewithlimits{argmin}_{i\in
\{1,...,M\}}E_{\b^{\prime }\mid \b,T_{i}}\left[ H(\b^{\prime })-H(\b)\right] \, 
\nonumber \\[-1ex]
&=&\,\operatornamewithlimits{argmin}_{i\in \{1,...,M\}}\left[
-\sum_{\b^{\prime }\in \mathbb{B}}\mathbf{Tr}_{\b\b^{\prime
}}(T_{i})\,\sum_{j=1}^{2^{M}}p_{j}^{\b^{\prime }}\log _{2}p_{j}^{\b^{\prime }}%
\right] \,,
\nonumber\\
&\left.\right.&
\end{eqnarray}%
where the second equality is obtained by removing constant terms.

\subsection{Dynamic Programming Entropy}

The Greedy-Entropy approach takes into account only the entropy in the next
step for selecting the experiment to be performed at the current step. If
the number $N$ of experiments is known a priori, then the dynamic
programming technique is used for finding an optimal entropy-based solution.
In~\citealp{huan2016sequential}, an approximate dynamic programming solution
based on the entropy scheme for cases with a continuous belief space is
provided. Here we employ the optimal dynamic programming solution, since the
belief space is finite. Again letting $\mu _{k}(\b):\b\rightarrow
\{T_{1},...,T_{M}\}$ be a policy at time step $k$, we define a bounded
immediate cost function at time step $k$ corresponding to transition from
the belief vector $\b(k)=\b$ to the belief vector $\b({k+1})=\b^{\prime }$ under
policy $\mu _{k}$ by 
\[
\tilde{g}_{k}(\b,\b^{\prime },\mu _{k}(\b))\,=\,H(\b^{\prime })\,-\,H(\b)\,,
\]%
for $k=0,...,N-1$. Define the terminal cost function by $\tilde{g}%
_{N}(\b)\,=\,H(\b)$, for any $\b\in \mathbb{B}$. Using the immediate and
terminal cost functions $\tilde{g}_{k}$ and $\tilde{g}_{N}$ instead of $g_{k}
$ and $g_{N}$, for $k=0,...,N-1$, in the dynamic programming process, the
optimal finite-horizon policy, $\mu _{0:N-1}^{\mathrm{Entropy}}(\b)$, for $%
\b\in \mathbb{B}$, based on the entropy, is obtained. It is called the
DP-Entropy policy.\vspace{0ex}

\section{Results}

\label{sec:NE}

\subsection{Simulation Set-Up}

According to the majority vote rule for generating Boolean network models of
gene regulatory networks, the $i$th Boolean predictor is given by 
\[
X_{i}(k+1)\,=\,f_{i}\left( X(k)\right) \,=\,%
\begin{cases} 1 & \text{If
}\sum_j R_{ij}\,X_j(k) >0,\\ 0 & \text{If }\sum_j R_{ij}\,X_j(k) <0,\\
X_i(k) & \text{If }\sum_j R_{ij}\,X_j(k) =0, \end{cases}
\]%
%
%
%
%
%
%
%
for $i=1,...,n$, where $R_{ij}$ can take three values: $R_{ij}=+1$ if there
is an activating regulation ($\mathcal{A}$) from gene $j$ to gene $i$, $%
R_{ij}=-1$ if there is suppressive regulation ($\mathcal{S}$) from gene $j$
to gene $i$, and $R_{ij}=0$ if gene $j$ is not an input to gene $i$ ~(%
\citealp{lau2007function}).

We employ the symmetric Dirichlet distribution for generating the initial
distribution over various network models 
\[
p^{\b}(0)\sim f(p^{\b}(0);\phi )=\frac{\Gamma \left( \phi \,2^{M}\right) }{%
\Gamma (\phi )^{2^{M}}}\,\prod_{j=1}^{2^{M}}p_{j}^{\b}(0)^{\phi -1}, 
\]%
%
%
%
%
%
%
%
where $\Gamma $ is the gamma function and $\phi >0$ is the parameter of the
symmetric Dirichlet distribution. The expected value of the initial
distribution for any value of $\phi $ is a vector of size $2^{M}$ with all
elements $1/2^{M}$. $\phi $ specifies the variability of the initial
distributions, the smaller $\phi $ is, the more the initial distributions
deviate from the uniform distribution.

\subsection{Performance Evaluation Based on Synthetic BNps}

To evaluate performance, simulations based on synthetic BNps have been
performed. 100 random BNps of size $6$ with a single set of $M$ unknown
regulations for each network have been considered. The perturbation
probability is set to $p=0.001$. 
Different values of $p$ have been tried and similar results, as presented in the sequel, have been observed. The states with up-regulated first gene are
assumed to be undesirable ($U=\{\mathbf{x}^{1},...,\mathbf{x}^{32}\}$).
Three different values are considered for the Dirichlet parameter: $\phi
=0.1,1$, and $10$. From each $\phi $, 500 initial distributions are
generated.

Five experimental design strategies are considered: 1) Greedy-MOCU, 2)
DP-MOCU, 3) Greedy-Entropy, 4) DP-Entropy, 5) Random. A successful
experimental design strategy has the ability to effectively reduce the cost
of intervention. Thus, the robust intervention based on the resulting belief
state of each strategy is applied to the true (unknown) network and the cost
of intervention (total steady-state mass in undesirable states) is used as a
metric. For a given belief state $\b(k)$ computed before taking the $k$th
experiment, the cost is $\xi _{\mathbf{\theta }^{\ast }}\left( \psi _{%
\mathrm{IBR}}^{\theta\mid\b(k)}\right) $. $H(\b(k))$ represents the amount of remaining
uncertainty in the system for a given belief state $\b(k)$. Thus, we define
the intervention gain of conducting the chosen experiment over a random
experiment by $\xi _{\mathbf{\theta }^{\ast }}\left( \psi _{\mathrm{IBR}%
}^{\theta\mid\b^{\mathrm{rnd}}(k)}\right) \,-\,\xi _{\mathbf{\theta }^{\ast }}\left(
\psi _{\mathrm{IBR}}^{\theta\mid\b(k)}\right) $, and the entropy gain as $H(\b^{\mathrm{%
rnd}}(k))\,-\,H(\b(k))$, where $\b(k)$ is the belief state after performing
the $k$th experiment determined via experimental design (Greedy-MOCU,
DP-MOCU, Greedy-Entropy or DP-Entropy), and $\b^{\mathrm{rnd}}(k)$ is the
belief vector before performing the $k$th experiment during the random
experimental design process.

\begin{figure*}[!tpb]
\centerline{\includegraphics[width=167mm]{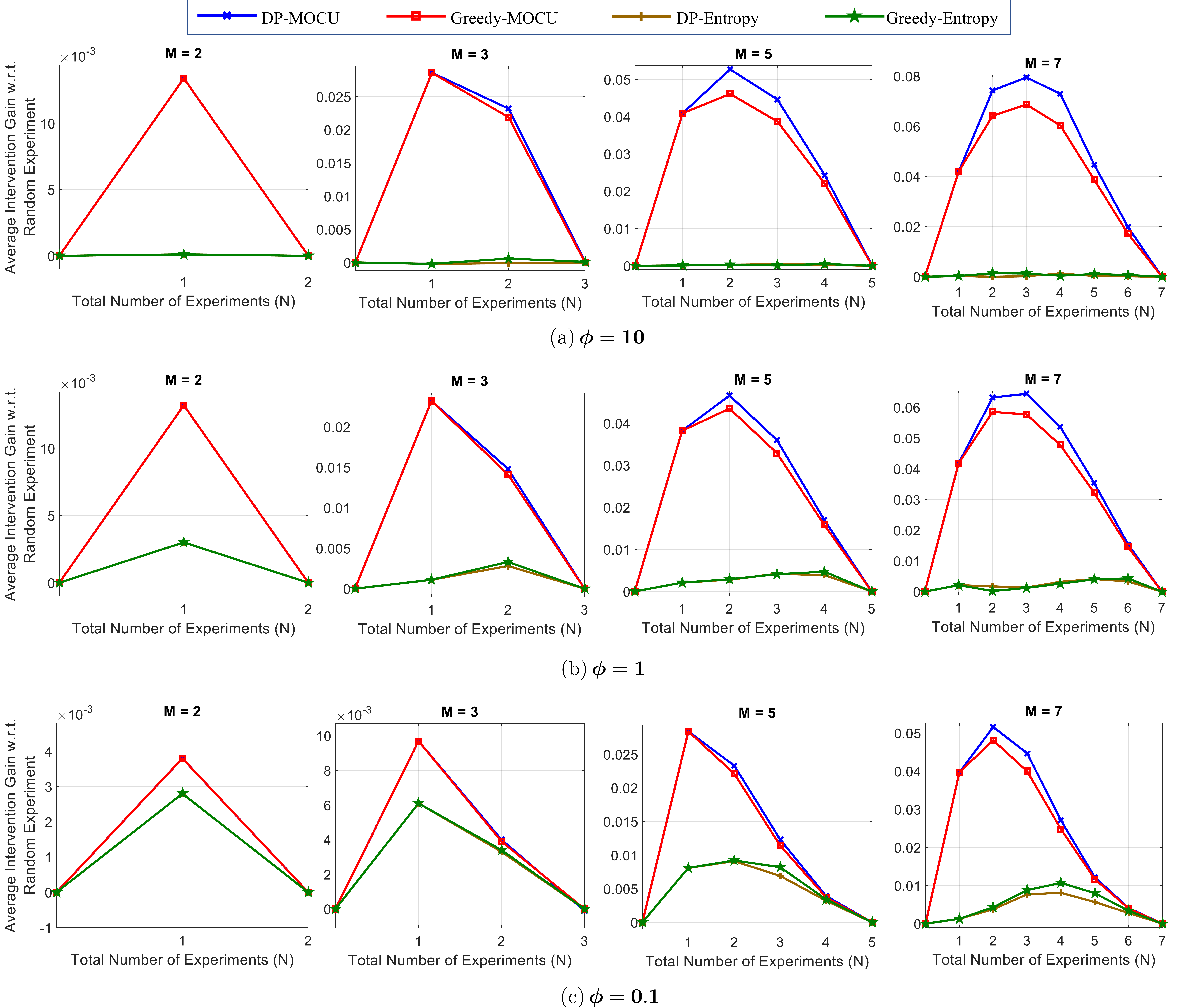}}\vspace{-2ex}
\caption{The average intervention gain with respect to the total number of
experiments, $N$, for randomly generated synthetic 6-gene Boolean networks
with 2, 3, 5, 7 uncertain regulations ($M$). }
\label{fig-Total-obj}
\end{figure*}

Figure~\ref{fig-Total-obj} shows the average gain of intervention with
respect to the horizon length $N$ strategies for different numbers of
unknown regulations ($M$) and Dirichlet parameters ($\phi$). {The figure
shows curves for $M=2,3,5,7$ and $\phi =0.1,1,10$}. 
The curves end at gain
value $0$ when $M=N$, so that all regulations have been identified. In
practice, the number of unknown regulations is usually more than the number
of experiments which can be performed. In these cases, large intervention
gains have been attained by the MOCU-based strategies in comparison to
entropy-based techniques, thus demonstrating the effectiveness of MOCU-based
strategies in reducing network uncertainty relevant to the objective.

The maximum amount of gain in MOCU-based strategies is achieved for $\phi
=10 $ (maximum uncertainty). In contrast, the intervention gains in
entropy-based strategies are very close to $0$ when the initial distribution
is closer to uniform ($\phi =10$) and increase as this distribution deviates
from uniform ($\phi =0.1$). Indeed, as $\phi $ gets larger (initial
distributions get closer to uniform), the Entropy scheme does not
discriminate between potential experiments and performs like a random
selection approach. Thus, entropy-based strategies slightly perform better
than the random strategy for non-uniform prior distributions, with their
performance being far worse than MOCU-based techniques. In addition, the
peak in the intervention gain is shifted slightly into the left side as $%
\phi $ decreases. This is due to the fact that in the presence of a
non-uniform initial distribution, the MOCU-based strategies are capable of
selecting the first most effective experiments in early steps to reduce the
intervention cost.

In Figure~\ref{fig-Total-obj}, DP-MOCU outperforms Greedy-MOCU in all cases
with respect to the cost of intervention, because future experiments are
taken into account for decision making in DP-MOCU, as opposed to
Greedy-MOCU, which only considers one-step look-ahead. When the total number
of experiments ($N$) is $1$, Greedy-MOCU and DP-MOCU are equivalent and the
same gain can be seen for both strategies. The highest gain difference is
achieved when the horizon length $N$ is less than $M$, the number of
uncertain parameters.

\vspace{0ex}

\begin{figure*}[h]
\centerline{\includegraphics[width=170mm]{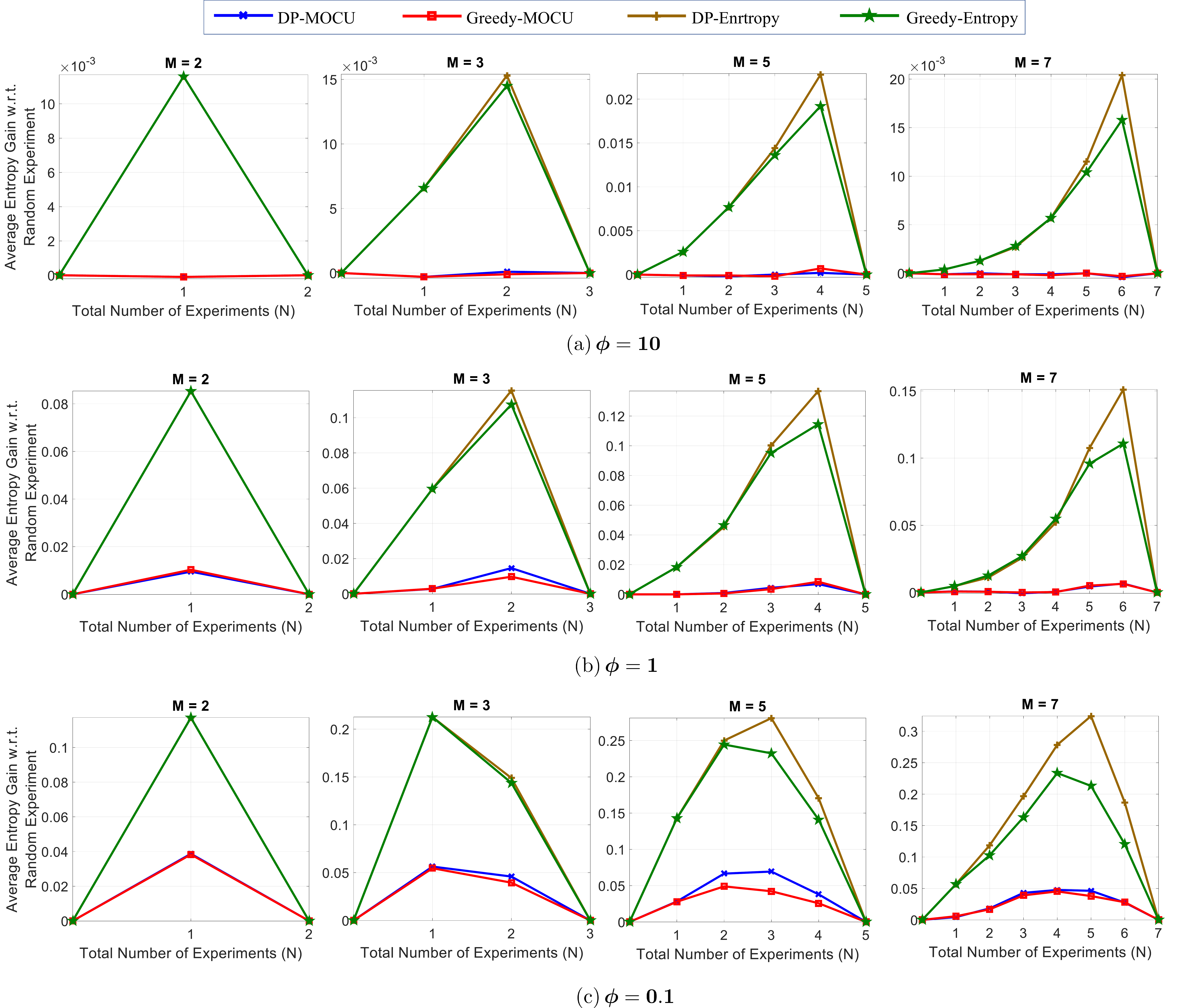}}\vspace{-2ex}
\caption{The average entropy gain with respect to the total number of
experiments, $N$, for randomly generated synthetic 6-gene Boolean networks
with 2, 3, 5, 7 uncertain regulations ($M$). }
\label{fig-Total-ent}
\end{figure*}
\vspace{0ex}

To better appreciate the performance of entropy-based techniques, the gain
in entropy is reported in Figure~\ref{fig-Total-ent}. Once again, {the figure
shows curves for $M=2,3,5,7$ and $\phi =0.1,1,10$.} 
Comparison between Figures~%
\ref{fig-Total-obj} and~\ref{fig-Total-ent} shows that the maximum entropy
reduction by the entropy-based strategies does not necessarily result in the
highest reduction in the cost of intervention, which is the main objective
of performing experimental design. Interestingly, while DP-Entropy is more
successful in reducing the entropy value in comparison to Greedy-Entropy (as
expected), it does not outperform Greedy-Entropy relative to average gain in
intervention.

Next, we consider the effect of the initial distribution on the performance
of various experimental design strategies. The horizon length and the number
of unknown regulations are set to be $N=4$ and $M=7,$ respectively. The
initial distribution is a vector of size $2^{7}$. The entropy of this
initial distribution specifies the amount of initial uncertainty in the
system. The closer this value is to its maximum value $7$, the closer the
initial distribution is to the uniform distribution. 
In the figure, we observe that as the entropy of the initial distribution
increases, the performance of both Greedy-MOCU and DP-MOCU increases as
well. This growth is higher for DP-MOCU compared to Greedy-MOCU, which shows
the superiority of DP-MOCU in reducing the intervention cost in the presence
of high uncertainty in the system. On the other hand, note the reduction
trend in the amount of intervention gain for the entropy-based techniques as
the entropy of the initial distribution increases. This is due to the fact
that the entropy-based strategies are unable to discriminate between
potential experiments in the presence of the uniform initial distribution
and perform like random selection.

\begin{figure}[!tpb]
\centerline{\includegraphics[width=85mm]{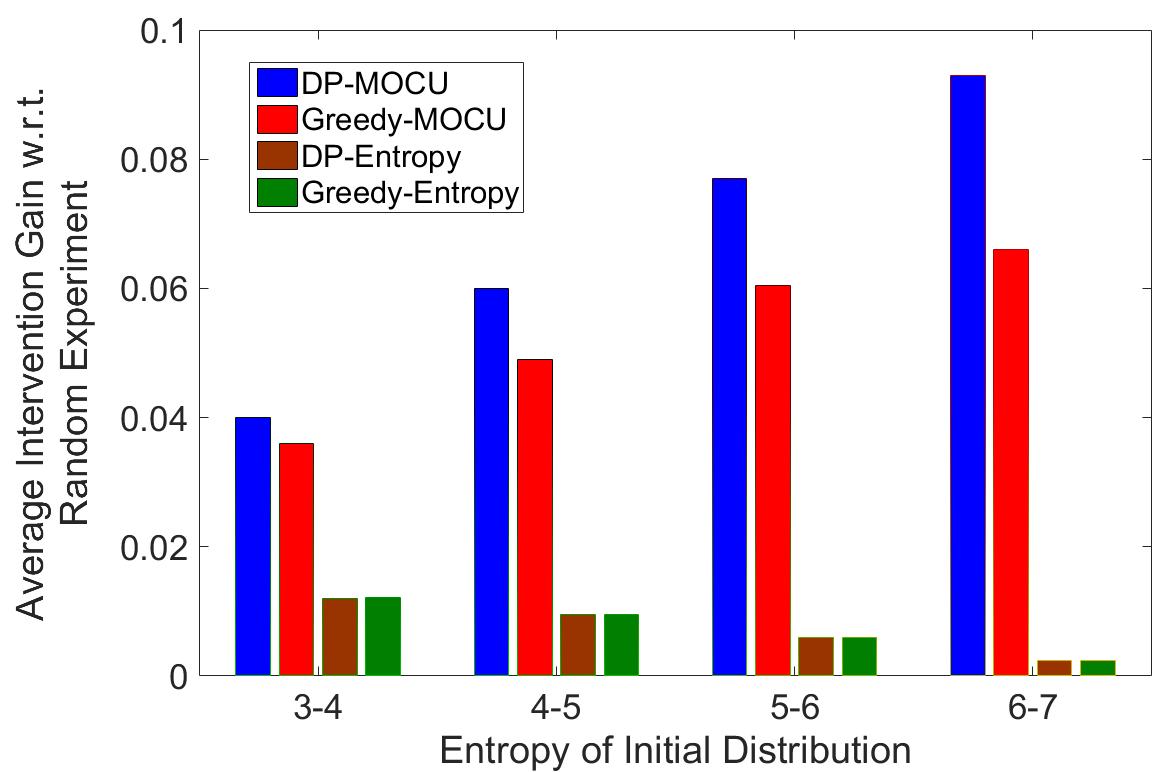}}\vspace{-2ex}
\caption{The average intervention gain with respect to the entropy of the
initial distribution, $p(0)$, for randomly generated synthetic 6-gene
Boolean networks with $7$ unknown regulations and total number of
experiments ($N$) equal to $4$.}
\label{fig-bar}
\end{figure}
\vspace{-0ex}

\begin{figure*}[!tpb]
\centerline{\includegraphics[width=170mm]{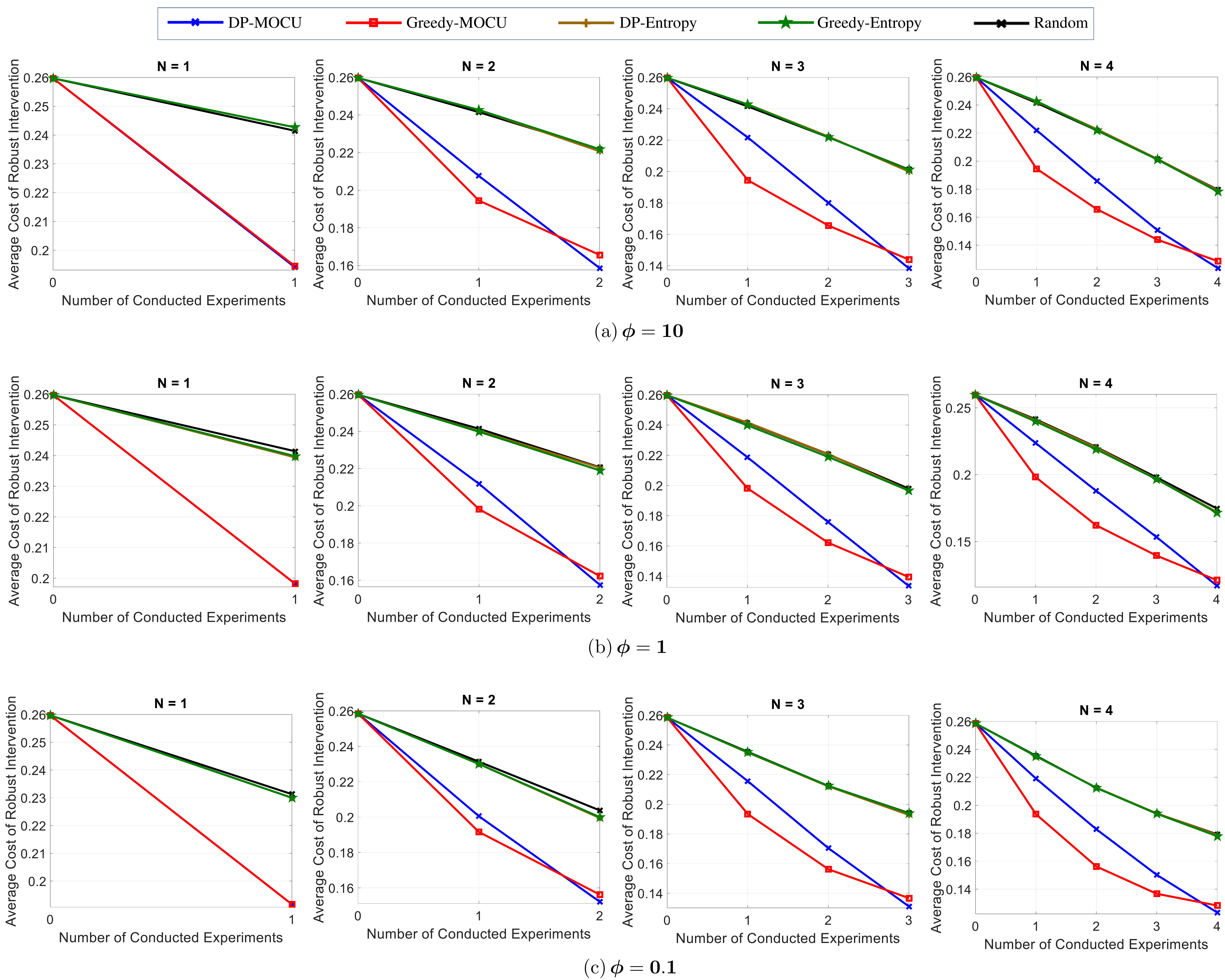}} \vspace{-2ex}
\caption{The average cost of robust intervention with respect to the number
of conducted experiments obtained by various experimental design strategies
for randomly generated synthetic BNps with $7$ unknown regulations ($M$) and 
{$\protect\phi=0.1,1,10$}.}
\label{fig-Condu-obj}
\end{figure*}

The average cost of robust intervention with respect to the number of
conducted experiments for different experimental design strategies is shown
in Figure~\ref{fig-Condu-obj}. 
%
%
%
DP-MOCU has the lowest average cost of robust intervention at the end of the
horizon (after taking all $N$ experiments); however, Greedy-MOCU has the
lowest cost before reaching the end of the horizon. This observation can be
understood by looking at the finite-horizon dynamic programming policy.
DP-MOCU finds a sequence of experiments from time $0$ to $N-1$ to minimize
the expected sum of the differences of MOCUs throughout this interval. The
expected value of MOCU after conducting the last experiment plays the key
role in the decision making by the dynamic programming policy. Thus, the
capability of DP-MOCU in planning for reducing MOCU at the end of the
horizon, as opposed to Greedy-MOCU which takes only the next step into
account for decision making, results in the lowest average cost of robust
intervention by DP-MOCU at the end of horizon. DP-MOCU and Greedy-MOCU are
equivalent for horizon length $N=1$ and behave differently for other cases.
When the number of experiments is not known a priori, Greedy-MOCU may be
preferred to DP-MOCU because, as presented in Figure~\ref{fig-Condu-obj},
the intervention gain in DP-MOCU might be lower than Greedy-MOCU before
conducting the total number of experiments.

\subsection{Performance Evaluation Based on the Mammalian Cell Cycle Network}

\begin{figure}[tbp]
\centerline{\includegraphics[width=70mm]{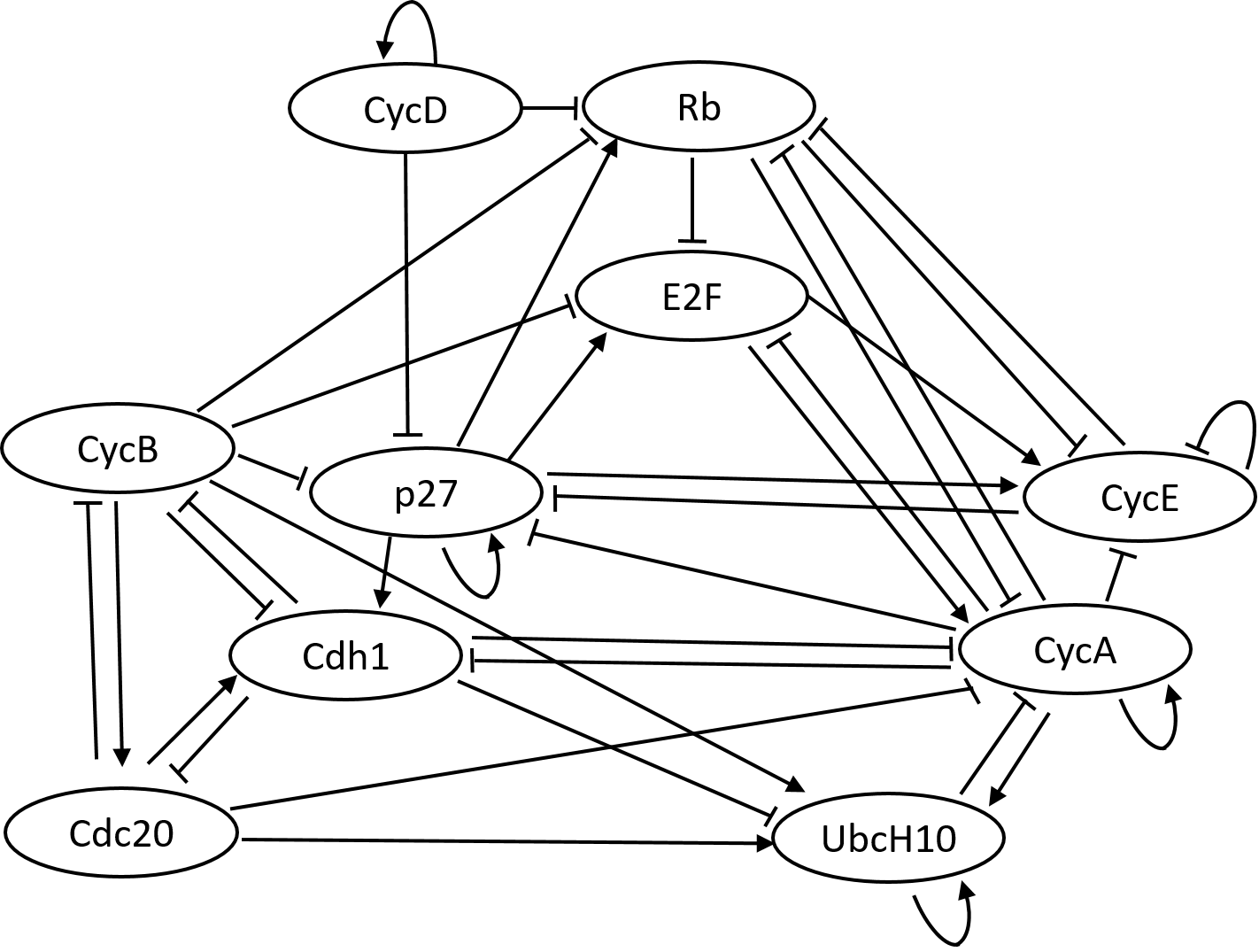}}\vspace{-2.2ex}
\caption{A gene regulatory network model of the mammalian cell cycle. Normal
arrows represent activating regulations and blunt arrows represent
suppressive regulations.}
\label{fig:mam}
\end{figure}

The mammalian cell cycle involves a sequence of events resulting in the
duplication and division of the cell. It occurs in response to growth
factors and under normal conditions, it is a tightly controlled process. A
regulatory model for the mammalian cell cycle, proposed in~%
\citealp{faure2006dynamical}, is shown in Figure~\ref{fig:mam}. This model
contains 10 genes: CycD, Rb, p27, E2F, CycE, CycA, Cdc20, Cdh1, UbcH10, and
CycB. The blunt and normal arrows represent suppressive ($\mathcal{S}$) and
activating ($\mathcal{A}$) regulations, respectively. Mammalian cell
division is coordinated with the overall growth of the organism via
extracellular signals that control the activation of CycD in the cell. Cell
division happens due to the positive stimuli activating Cyclin D (CycD).
When CycD is up-regulated, it inactivates the tumor suppressor Rb protein
via phosphorylation. Rb can also be expressed if gene p27 and either CycE or
CycA are active. The activation of Rb in the absence of stimuli causes
cell-proliferative (cancerous) phenotypes. States with down-regulated CycD,
Rb, and p27 $(X_{1}=X_{2}=X_{3}=0)$ are undesirable, representing cancerous
phenotypes. The goal is to reduce the steady-state probability mass of the
set of undesirable states, $U=\{\mathbf{x}^{1},...,\mathbf{x}^{128}\}$, via
structural intervention. 

We consider various cases with $2$ to $6$ unknown regulations ($M$). We
randomly select $100$ different sets of $M$ regulations from the network,
for which we assume their regulatory information is not known, and apply
various experimental design strategies to predict the experiment to be
performed. 500 initial distributions have been generated from the Dirichlet
distribution with parameter $\phi =1$.

\begin{figure*}[tbp]
\centerline{\includegraphics[width=170mm]{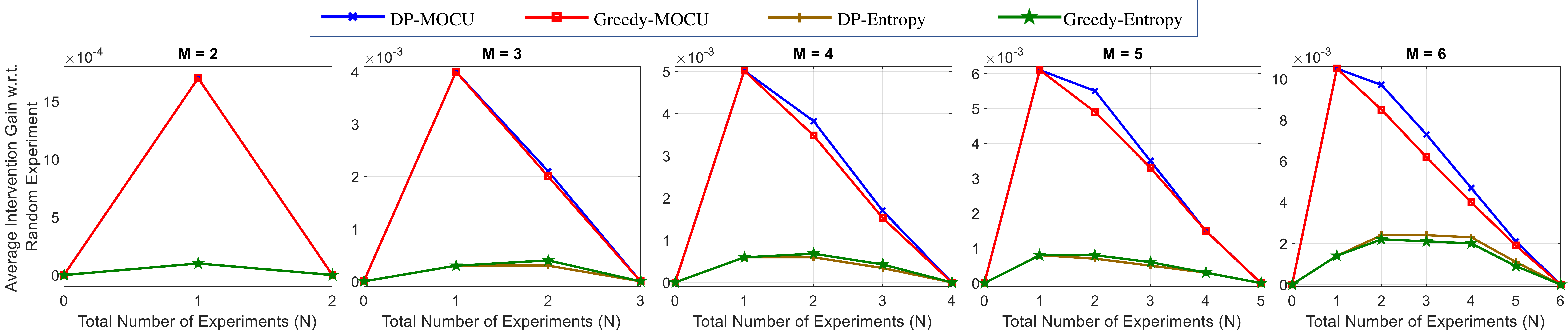}}\vspace{-1ex}
\caption{The average intervention gains of various experimental strategies
versus the random strategy for the Mammalian Cell Cycle network for 2 to 6 unknown regulations ($M$) and $\protect\phi=1$.}
\label{fig-Mam-Total-obj-ent}
\end{figure*}

The average intervention gains for various experimental design strategies
are presented in Figure~\ref{fig-Mam-Total-obj-ent}, which shows curves for {$%
M=2,...,6$.} 
DP-MOCU
and Greedy-MOCU have the highest average intervention gain in comparison to
the entropy-based strategies. DP-MOCU is clearly superior to Greedy-MOCU for
cases with larger numbers of unknown regulations and when the number of
experiments is smaller than the number of unknown regulations ($1<N<M$).
Both Greedy-Entropy and DP-Entropy perform poorly in all cases. 

\subsection{Computational Complexity Analysis}

Consider a network with $n$ genes, in which the states $2^{n-1}$ to $2^{n}$
are undesirable. Structural intervention requires $2^{n-1}$ searches over $%
2^{n}\times 2^{n}$ state pairs. This gives complexity $O(2^{3n})$ for the
optimal intervention process for a single network. Given $M$ uncertain
parameters, which poses $2^{M}$ different network models, the complexity of
Greedy-MOCU is of order $O(2^{M}\times 2^{3n})$. On the other hand, DP-MOCU
has an extra step for the dynamic programming process. The complexity of the
dynamic programming process is of order $O(3^{2M}\,\times N)$, where $N$ is
the horizon length. Thus, the complexity of DP-MOCU is $O(\max
\{3^{2M}\,\times N,2^{M}\times 2^{3n}\})$. In contrast to the MOCU-based
strategies, the complexities of the entropy-based techniques are independent
of the intervention process. Greedy-Entropy and DP-Entropy have complexities 
$O(2^{M}\times 2^{n})$ and $O(\max \{3^{2M}\,\times N,2^{M}\times 2^{n}\})$,
respectively.

Table~\ref{table:cost} shows approximate processing times for networks of
different size with various numbers of regulations. Simulations have been
run on a machine with 16 GB of RAM and Intel(R) Core(TM) i7 CPU, 3.6 GHz.
The running time of the MOCU-based strategies grows exponentially as the
number of genes increases. It also increases with increases in the number of
unknown regulations. It can be seen that the running time of DP-MOCU is
slightly higher than that of Greedy-MOCU owing to the extra dynamic
programming recursion in DP-MOCU.

Clearly, computational complexity is an issue. The issue has been addressed
in the context of structural intervention in~%
\citealp{dehghannasiri2015efficient}, where computation reduction for
MOCU-based design is achieved via network reduction schemes. These result in
suboptimal experimental design, but they are still superior to random design.

\begin{table}[ht!]
\caption{Comparing the approximate processing times (in seconds) of various
experimental design methods for networks of size $n$ with $M$ uncertain
regulations, and $N=3$}
\label{table:cost}{\scriptsize \centering
\begin{tabular}{ccccc}
\cmidrule{3-5} \cmidrule{3-5} &  & $n=10$ & $n=11$ & $n=12$ \\[0pt] 
\midrule & $M=4$ & 250 & 2651 & 32933 \\[0pt] 
\cmidrule{2-5} Greedy-MOCU & $M=5$ & 493 & 5210 & 65208 \\[0pt] 
\cmidrule{2-5} & $M=6$ & 967 & 10264 & 127397 \\[0pt] 
\midrule & $M=4$ & 272 & 2696 & 32989 \\[0pt] 
\cmidrule{2-5} DP-MOCU & $M=5$ & 490 & 5294 & 65323 \\[0pt] 
\cmidrule{2-5} & $M=6$ & 1002 & 10314 & 127413 \\[0pt] 
\midrule & $M=4$ & 5 & 11 & 21 \\[1pt] 
\cmidrule{2-5} DP-Entropy & $M=5$ & 15 & 29 & 63 \\[1pt] 
\cmidrule{2-5} & $M=6$ & 44 & 86 & 173 \\[0pt] 
\midrule & $M=4$ & 6 & 13 & 25 \\[1pt] 
\cmidrule{2-5} Greedy-Entropy & $M=5$ & 18 & 36 & 69 \\[1pt] 
\cmidrule{2-5} & $M=6$ & 50 & 99 & 184 \\[0pt] 
\midrule &  &  &  & 
\end{tabular}%
}
\end{table}
\vspace{0ex}

\section{Conclusion}

By taking into account the operational objective, MOCU-based experimental
design significantly outperforms entropy-based design. Our aim in this paper
has been twofold: to demonstrate this advantage and to propose and examine
the effect of using finite-horizon dynamic programming for sequential
design. The simulations show that if one has a fixed number of experiments
in mind, then dynamic programming provides improved results because it takes
into account experiments over the full horizon, but for the same reason it
can be disadvantageous if one is interested in stopping experimentation once
MOCU reduction falls below a given threshold, meaning that further
experimentation is not worth the cost. While our focus in this paper has
been intervention in gene regulatory networks, it should be recognized that
Greedy-MOCU has been used for optimal experimental design in other
environments such as the development of optimally performing materials (%
\citealp{dehghannasiri2017optimal}) and optimal signal filtering (%
\citealp{dehghannasiri2017IET}). Dynamic programming can be applied for
these problems in the same way it has been done here.\vspace{0ex}

\begin{acks}
The authors acknowledge the support of the National Science Foundation, through NSF award CCF-1718924.
\end{acks}

\end{document}